\documentclass[aps,prb,twocolumn,showpacs,superscriptaddress,10pt,reprint]{revtex4}

\usepackage{graphicx}
\usepackage{amsmath,amssymb,amstext,amsthm}
\usepackage{bbold}
\usepackage{cancel}
\usepackage{color}
\usepackage{verbatim}
\usepackage{booktabs}
\usepackage[caption=false]{subfig}

\newcommand{\Z}{\mathbb{Z}}
\newcommand{\id}{\mathbb{1}}

\usepackage[utf8x]{inputenc}

\begin{document}

\title{One-dimensional Dirac electrons on the surface of weak topological 
insulators}

\author{Alexander Lau}
\affiliation{Institute for Theoretical Solid State Physics, IFW Dresden, 
01171 Dresden, Germany}

\author{Carmine Ortix}
\affiliation{Institute for Theoretical Solid State Physics, IFW Dresden, 
01171 Dresden, Germany}

\author{Jeroen van den Brink}
\affiliation{Institute for Theoretical Solid State Physics, IFW Dresden, 
01171 Dresden, Germany}
\affiliation{Department of Physics, TU Dresden, 01062 Dresden, Germany}

\date{\today}

\begin{abstract}

We show that a class of weak three-dimensional topological insulators feature 
one-dimensional Dirac electrons on their surfaces. 
Their hallmark is a line-like energy dispersion along certain directions of the 
surface Brillouin zone. 
These one-dimensional Dirac line degeneracies are topologically protected by a
symmetry that we refer to as {\it in-plane time-reversal invariance}. 
We show how this invariance leads to Dirac lines in the surface spectrum of 
stacked Kane-Mele systems and more general models for weak three-dimensional 
topological insulators.
 
\end{abstract}

\pacs{
73.20.-r, 
73.43.-f, 
72.25.-b, 
85.75.-d  
}

\maketitle


\section{Introduction}
\label{sec:intro}

The field of topological properties of matter is one of the most
active areas in condensed matter physics.~\cite{QiZ11,Fu11,LaT13,LaT14,AFG14}
In particular, topological insulators 
(TIs) have attracted a lot of attention in recent years due to their unique 
physical properties and potential applications.~\cite{HaK10,Moo10}
In contrast to conventional 
insulators, TIs feature topologically robust gapless edge or surface states with
a linear Dirac
dispersion. Most interestingly, these
so-called topological surface states are a direct physical consequence of 
the nontrivial topology of the bulk band structure and protected by 
time-reversal symmetry. 

The number of inequivalent classes of TIs depends on the dimension 
of the systems under consideration.~\cite{FuK07,FKM07}
In two dimensions, there is only one class which
is characterized by a single topological invariant. Experimental realizations
are, for instance, HgTe quantum well structures.~\cite{BHZ06,KWB07}
In three dimensions,
TIs are described by a quadruple of topological invariants and their 
variety is, therefore, much larger. 
One then distinguishes between weak and strong TIs. 
The latter possess an odd number of Dirac-like surface states at the 
time-reversal
invariant (TRI) momenta of the surface Brillouin zone (BZ). 
For this reason, they are stable against 
generic
perturbations preserving time-reversal symmetry.
On the contrary, weak TIs are unstable against 
translational-symmetry breaking perturbations, such as charge density waves, 
due to the presence of an even number of Dirac-like
degeneracies in the surface band structure. Furthermore, they are topologically 
equivalent to a stack of
layers of two-dimensional (2D) TIs. Both classes have been identified 
experimentally,
including, for instance, the strong TIs 
Bi$_{1-x}$Sb$_x$,~\cite{BHZ06,KWB07,HXW09}
Bi$_2$Te$_3$,~\cite{CAC09} Bi$_2$Se$_3$,~\cite{ZLQ09}
as well as the recently found weak TI  Bi$_{14}$Rh$_3$I$_9$.~\cite{RIR13}

The topologically protected surface Kramers doublets
found in three-dimensional (3D) TIs are usually referred
to as Dirac points,~\cite{FuK07,LQZ12}
since in their vicinity 
the dispersion of the surface states
resembles that of a 2D Dirac electron (see Fig.~\ref{fig:in_plane_TRS}(b)). 
However, a recent
experimental work by Gibson \textit{et al.}~\cite{GEY14} suggests that
3D TIs can also exhibit Dirac-like {\it line} degeneracies 
on their surfaces corresponding to effectively one-dimensional (1D) Dirac 
electrons. 
By angle-resolved photoemission spectroscopy they established the presence of 
highly anisotropic
surface states on the (113) surface of Ru$_2$Sn$_3$, which show an almost
line-like dispersion along certain high-symmetry directions. 

Thus far, 1D Dirac states have not been investigated in the context of 3D 
topological insulators.
The first question that arises is whether and under what circumstances 
topological
insulators can exhibit Dirac-like line degeneracies on their surface, and, 
secondly, whether 
they are protected by any kind of symmetry. In this paper, 
we establish that Dirac lines naturally occur in weak topological 
insulators during the transition from effectively two dimensions to three 
dimensions. Furthermore, these lines are topologically protected by a symmetry 
that we refer to as \emph{in-plane time-reversal invariance}. 

The paper is organized as follows: in Sec.~\ref{sec:in_plane_tri} we introduce 
the concept of in-plane time-reversal invariance, thereby showing how it leads 
to topologically protected Dirac lines on the surface of weak TIs. 
In Secs.~\ref{sec:stacked_KM_model},~\ref{sec:cubic_LQZ_model} 
we will explicitly demonstrate how Dirac lines arise due to in-plane time-reversal 
invariance in two paradigmatic weak TI model Hamiltonians, namely stacked Kane-Mele 
systems,~\cite{KaM05_1,KaM05_2} and the cubic Liu-Qi-Zhang model.~\cite{LQZ12}
Finally we will draw our conclusions in Sec.~\ref{sec:conclusions}.


\section{In-plane time-reversal invariance}
\label{sec:in_plane_tri}

Let us consider a 
stack of arbitrary but identical 2D TIs
where the stacking direction is,
without loss of generality, the $z$ direction. Furthermore, let the system be 
finite in a direction perpendicular to the stacking direction, e.g. the $x$
direction. Generically, 
each of the 2D TI ribbons
will have topologically protected
edge states inside the bulk energy gap with an edge Kramers doublet at a TRI 
point.~\cite{KaM05_2,FuK06}
Let us first inspect the case where the layers are completely decoupled and 
the band structure of the
system does not disperse in the $k_z$ direction. Thus, 
the 
Kramers doublet of the 1D BZ
of the layers become perfectly flat line degeneracies, more precisely 
Dirac lines, in the surface BZ of the stack.
With this stacking procedure we have constructed
a weak topological insulator with $\Z_2$ 
indices $(\nu_0;\nu_1\nu_2\nu_3)=(0;001)$,~\cite{FKM07,FuK07}
and with Dirac lines
instead of Dirac points on its surfaces. This is not surprising, since the 
stack of 
2D TIs is still an effectively 2D system.
Generally, one would expect the Dirac lines to break up and leave only Dirac 
points at TRI momenta once an arbitrary time-reversal interlayer coupling is 
introduced.
This, in turn, leads to a typical surface band dispersion of a weak TI with an 
even number of surface Dirac cones.~\cite{FKM07}
However, in the following we will show that this is not necessarily the case
and that the presence of in-plane time-reversal invariance leads to
topologically protected Dirac lines even in the full 3D system with coupled 
layers.

\begin{figure}[t]\centering
\includegraphics[width=0.9\linewidth] {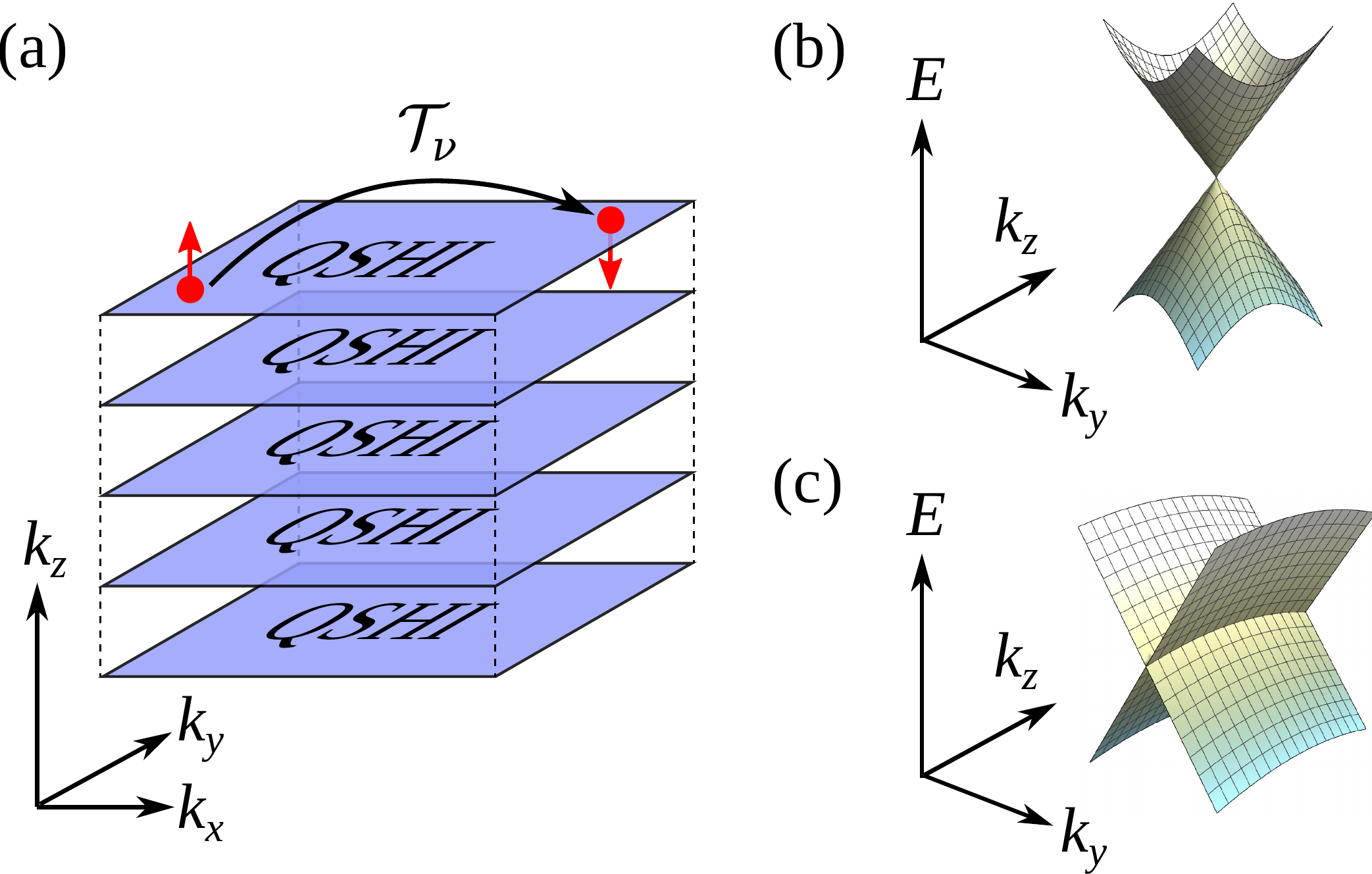}
\caption{(color online). (a) Illustration of the in-plane time-reversal
operator $\mathcal{T}_\nu$ in momentum space: the operator flips the 
spin of an electron and reverses the momentum component parallel to the 
plane defined by the weak indices $\nu=(\nu_1,\nu_2,\nu_3)$. 
As demonstrated, a weak TI with $\mathcal{T}_\nu$ symmetry 
yields a QSHI for any fixed $k_z$ which gives rise to topologically
protected Dirac edge states associated with each layer. (b) Typical 
dispersion of a 2D Dirac particle on the surface of a TI. (c) Typical
dispersion of a 1D Dirac particle on the surface of a weak TI
with $\mathcal{T}_\nu$ symmetry.}
\label{fig:in_plane_TRS}
\end{figure} 

In-plane time-reversal invariance is the 2D analogue of the \emph{conventional} 
time-reversal symmetry with respect to a
specific plane $\Pi$ [see Fig.~\ref{fig:in_plane_TRS}(a)]. 
The corresponding antiunitary operator 
$\mathcal{T}_\Pi$ acts 
similarly to its 3D analogue except that the momentum component 
perpendicular to the plane $\Pi$ remains unchanged. 
Without loss of generality, we choose the plane to be $(001)$. Then,
the in-plane time-reversal operator for spin-$\frac{1}{2}$ particles reads
\begin{equation}
\mathcal{T}_{(001)} := i(\id \otimes s^y)K\:\:
\mathrm{with}\:\: k_x,k_y,k_z\rightarrow -k_x,-k_y,k_z.
\end{equation}
Here, $s^y$ acts on the spin part of the system, where, again without loss of 
generality, we have chosen $z$ as the quantization axis of spin and the Pauli 
matrices
as a spin representation. The identity matrix $\id$ is 
in the space 
spanned by additional degrees of freedom, such as orbital or sublattice. 
We further denote the matrix 
part of the operator,
without the action on momentum, by $T_{(001)}$. Note that $\mathcal{T}_{(001)}$ 
has the same 
operator structure as the conventional time-reversal operator $\mathcal{T}$.
Obviously, for a 2D system $\mathcal{T}_{(001)}$ 
and $\mathcal{T}$ are identical. 
If we deal with a weak TI with weak indices
$\nu=(\nu_1,\nu_2,\nu_3)$, we denote the in-plane time-reversal operator 
associated with the $(\nu_1,\nu_2,\nu_3)$ plane by $T_\nu$.  

Since both types of time-reversal operators appear to be very similar, one 
expects the presence of an analogue 
of Kramers theorem for the in-plane 
time-reversal symmetric system. To establish this, we recall that a 
time-reversal
symmetric system satisfies 
$T H(k_x,k_y,k_z)T^{-1} = H(-k_x,-k_y,-k_z)$, i.e. 
the Bloch Hamiltonian commutes with $T$ at TRI points 
where $(k_x,k_y,k_z)=(-k_x,-k_y,-k_z)$. 
For particles with half-integer spin, this implies Kramers theorem.~\cite{HaK10}
For an in-plane time-reversal symmetric system, we have instead
\begin{equation}
T_{(001)} H(k_x,k_y,k_z)T_{(001)}^{-1} = H(-k_x,-k_y,k_z).
\label{eq:T_nu_symmetry_condition}
\end{equation}
Hence, for half-integer particles the energy spectrum must be
at least two-fold degenerate along
\emph{in-plane time-reversal invariant lines}
satisfying $(k_x,k_y,k_z)=(-k_x,-k_y,k_z)$. 
This thus gives rise to \emph{Kramers lines}
along which Kramers-like degeneracies are
guaranteed and topologically protected by $\mathcal{T}_{(001)}$ invariance.
Since
this condition includes also the 8 TRI points in the BZ,
the Kramers lines always connect certain Kramers points in the BZ.

By the bulk-edge correspondence, one generically finds topologically protected 
Dirac lines on surfaces perpendicular to the stacking direction.
The Dirac lines resemble the dispersion of a 1D Dirac 
particle in a 2D space: along one direction, it shows the typical linear
Dirac dispersion whereas the dispersion is line-like in the 
perpendicular direction. This is illustrated in Fig.~\ref{fig:in_plane_TRS}(c).
If $\mathcal{T}_\nu$ is broken without breaking
conventional time-reversal symmetry, only degeneracies at the Kramers points
in the BZ are still topologically protected. Therefore, each Dirac line splits
and the associated 1D Dirac particle decays into two 2D Dirac particles. 

Precisely for this reason, topologically protected Dirac lines cannot exist in 
strong TIs.
Consider a strong TI with one Dirac point and without any in-plane time-reversal
symmetry. It is always possible to reduce the 
number of Dirac points of a strong TI to one by introducing suitable 
translational symmetry breaking and surface potential terms. Now choose an 
arbitrary
plane $(klm)$ and establish the corresponding $\mathcal{T}_{klm}$ symmetry. 
Before that, the single Dirac point of the strong TI must be connected 
to the bulk continuum along any line that will later become a Kramers
line once in-plane time-reversal symmetry has been established. 
Otherwise, there would be a second Dirac point at the opposite 
TRI momentum and the system would not be a strong TI. In the process
of establishing $\mathcal{T}_{klm}$ symmetry, the ``arms'' of the Dirac point, 
therefore, 
pull down the bulk bands to which they are attached and thereby close the
bulk energy gap. 
Thus the system cannot be an insulator and either becomes a semimetal 
or a metal.
For this reason, the presence of in-plane time-reversal
symmetry, which is essential for the appearance of Dirac lines, is not 
reconcilable
with a strong TI. A demonstration of this 
feature will be explicitly shown in Sec.~\ref{sec:cubic_LQZ_model}.

There is another appealing point of view which illustrates the connection
between in-plane time-reversal symmetry and Dirac lines in weak TIs.
For this, we consider a weak TI with
weak indices (001) such that the associated stacking direction 
is the $z$ direction.
Furthermore, let us treat $k_z$ in the corresponding 
Hamiltonian $H(k_x,k_y;k_z)$
as a parameter. If now in-plane time-reversal symmetry $\mathcal{T}_{(001)}$ 
is present
in the weak TI, we have a 2D quantum spin-Hall insulator (QSHI) for any value 
of the 
parameter $k_z$ [see Fig.~\ref{fig:in_plane_TRS}(a)]. This can be easily seen. 
First, 
this is clearly the case for $k_z=0$ and $k_z=\pi$ since the system
obeys conventional time-reversal symmetry and is topologically non-trivial. 
However,
as we move along $k_z$ the 2D systems always preserve time-reversal symmetry
in the 2D sense due to the presence of in-plane time-reversal symmetry in the 
full
3D system. Moreover, the bulk gap does not close. Therefore, we cannot have a 
topological phase transition and the collection of 2D systems stay in the
QSHI phase for
all $k_z$. This also implies that any such 2D QSHI will have topologically
protected (spin-filtered) edge states that will form edge Kramers doublets at
the TRI momenta of the 1D
BZ. When we now move along $k_z$, the edge Kramers
doublets cannot be broken, which implies the existence of Dirac lines along 
the in-plane TRI lines in surface BZ of the weak TI.

On top of this, one can associate a "line" of topological invariants with an 
insulator respecting in-plane time-reversal symmetry. Indeed, for each value 
of the stacking direction momentum $k_z$ we can define a 2D topological $\Z_2$ 
invariant $\nu(k_z)$,~\cite{FuK06} which cannot change as we move along $k_z$. 
Hence, the line of topological invariants can only assume two constant values:
$\nu(k_z)=0$ for a trivial insulator, and $\nu(k_z)=1$
for an in-plane time-reversal invariant weak TI hosting 1D Dirac electrons 
on its surfaces.
         

\section{Stacked Kane-Mele model}
\label{sec:stacked_KM_model}

Having established the general consequences of in-plane time reversal 
invariance, we apply these notions explicitly to stacked Kane-Mele systems.  
The Kane-Mele model~\cite{KaM05_1,KaM05_2} 
is known to be a realization of a 
2D TI in certain parameter ranges. It comprises a 
2D nearest-neighbor tight-binding model on a honeycomb lattice 
with additional $\mathcal{T}$-invariant spin-orbit interaction terms, 
where $\mathcal{T}$ is the \emph{conventional} time-reversal operator
(see Ref.~\onlinecite{HaK10}). The model is described by the following 
Hamiltonian~\cite{KaM05_1,KaM05_2}
\begin{eqnarray}
\mathcal{H}_\textrm{KM} &&= t\sum_{\langle\mathbf{i},\mathbf{j}\rangle,\sigma}
c_{\mathbf{i}\sigma}^\dagger c_{\mathbf{j}\sigma}
+ i\lambda_\mathrm{SO} 
\sum_{\langle\langle\mathbf{i},\mathbf{j}\rangle\rangle, \sigma\sigma'}
 \nu_{\mathbf{i}\mathbf{j}}\, c_{\mathbf{i}\sigma}^\dagger 
s_{\sigma\sigma'}^z c_{\mathbf{j}\sigma'}\nonumber\\
&&{}+ i\lambda_\mathrm{R} 
\sum_{\langle\mathbf{i},\mathbf{j}\rangle, \sigma\sigma'}
 c_{\mathbf{i}\sigma}^\dagger 
(\mathbf{s} \times \hat{\mathbf{d}}_{\mathbf{i}\mathbf{j}})^z_{\sigma\sigma'}
 c_{\mathbf{j}\sigma'}
+ \lambda_\nu \sum_{\mathbf{i}\sigma} \xi_\mathbf{i}\, 
c_{\mathbf{i}\sigma}^\dagger c_{\mathbf{i}\sigma}, \nonumber
\label{eq:KM_hamiltonian}
\end{eqnarray}
where the notations and coefficients are as in Ref.~\onlinecite{KaM05_2}.
We note that the second term describes in-plane
$z\rightarrow -z$ symmetric spin-orbit
coupling (SOC), the third term represents in-plane Rashba SOC, and 
the third term is a staggered sublattice potential which breaks inversion
symmetry in the plane. 
As usual, we denote the two interpenetrating hexagonal sublattices of the 
honeycomb lattice as $A$ and $B$ and follow the conventions of 
Ref.~\onlinecite{KaM05_2}.

From this 2D model we build up a 3D system by
stacking the Kane-Mele layers along the $z$ direction. This is done in such a 
way
that corresponding lattice points of the same sublattice 
but in different layers lie
on top of each other (AA stacking). In order to couple the layers, we 
introduce a nearest-neighbor interlayer hopping term and an 
interlayer SOC term, leading us to the model
Hamiltonian
\begin{eqnarray}
\mathcal{H} &=& \sum_l \mathcal{H}_{\textrm{KM},l} + 
\tau \sum_{\langle l,l'\rangle}
\sum_{\mathbf{i}\sigma} c_{\mathbf{i}l\sigma}^\dagger c_{\mathbf{i}l'\sigma}
\nonumber\\
&&{}+ i\lambda_{\textrm{SO},\perp} \sum_{\langle l,l'\rangle}
\sum_{\mathbf{i}\sigma\sigma'} \mu_{ll'}\, c_{\mathbf{i}l\sigma}^\dagger 
s_{\sigma\sigma'}^z c_{\mathbf{i}l'\sigma'},
\end{eqnarray}
where $l$ indexes the layers, and $\mu_{ll'}=\pm1$ for $l\gtrless l'$. 
We note that after a Fourier transformation, the corresponding Bloch
Hamiltonian $H(\mathbf{k})$ is a $4\times 4$ matrix which 
can be expanded in terms of Dirac matrices and their commutators
similar to Ref.~\onlinecite{KaM05_2}.
Furthermore, these matrices can be written as Kronecker products
of Pauli matrices $\tau^i$ in sublattice space and Pauli matrices $s^i$
in spin space. In this way, the interlayer terms of the Hamiltonian become
\begin{eqnarray}
H_{ih}(\mathbf{k}) &=& 2\tau\cos k_z\,(\id\otimes\id),
\label{eq:interlayer_hopping}\\
H_{\mathrm{SO},\perp}(\mathbf{k}) &=& -2\lambda_{\textrm{SO},\perp}
\sin k_z\,(\id\otimes s^z),
\label{eq:interlayer_soc}
\end{eqnarray}
where the interlayer distance has been set to unity. The other terms of the
Hamiltonian can be found in Ref.~\onlinecite{KaM05_2}. The relevant in-plane
time-reversal operator for this model is 
$\mathcal{T}_{(001)}=i(\id \otimes s^y)K$ with
$(k_x,k_y,k_z)\rightarrow (-k_x,-k_y,k_z)$.

In particular, we are interested in surface states
which will be studied for a slab of thickness $W$ with $(010)$
surfaces, where $W$ is measured in the number of unit cells. 
The surface cuts out so-called zigzag edges from
each Kane-Mele layer. In other words, we will investigate a stack of
infinitely many Kane-Mele layers with zigzag termination. 
The corresponding Bloch Hamiltonian $H^{(010)}(k_x,k_z)$
of the slab is a $4W\times 4W$ matrix whose
energies are obtained by exact diagonalization.

\begin{figure}[t]\centering
\subfloat{\includegraphics[width=0.49\linewidth]
{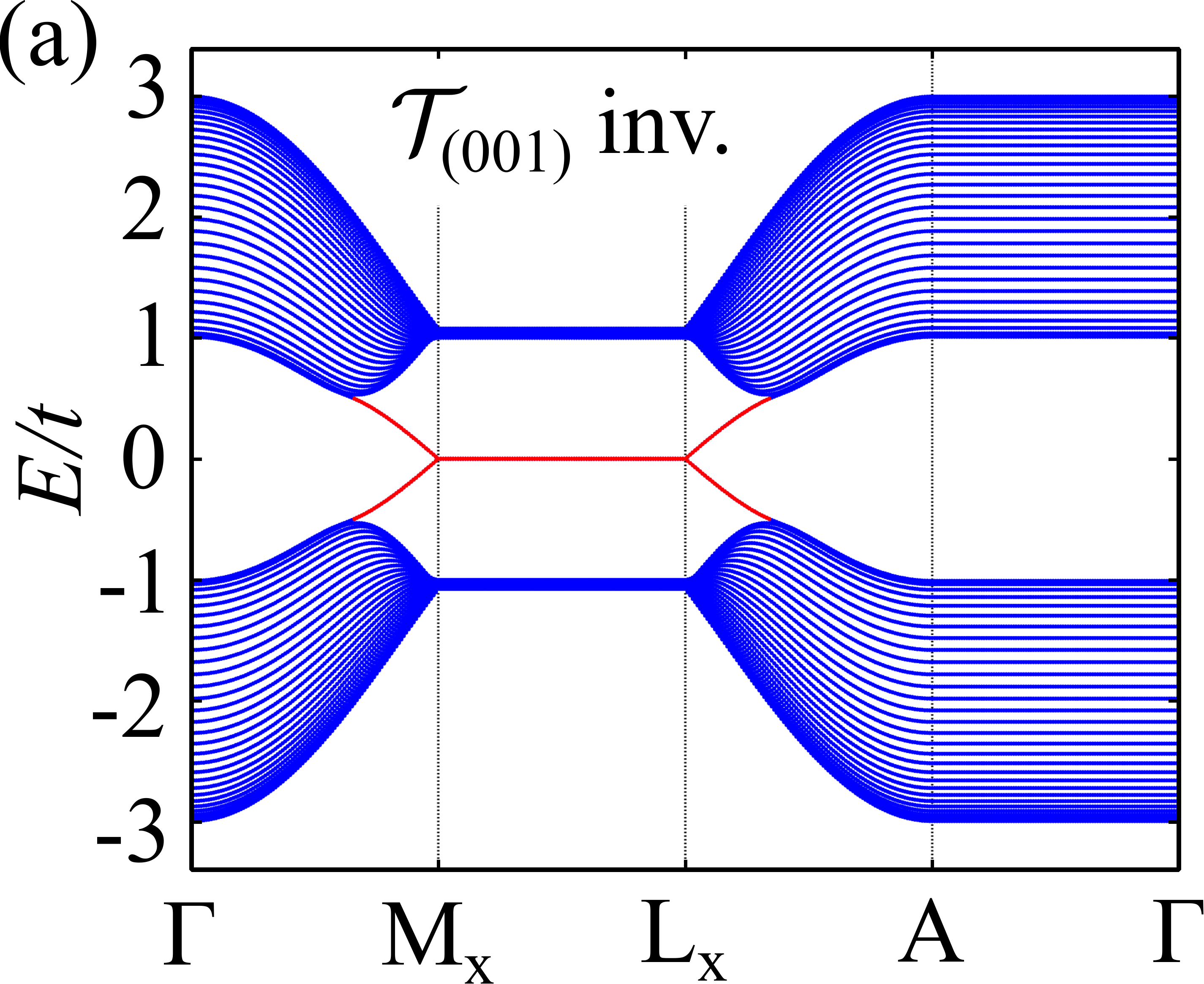}}
\hfill
\subfloat{\includegraphics[width=0.49\linewidth]
{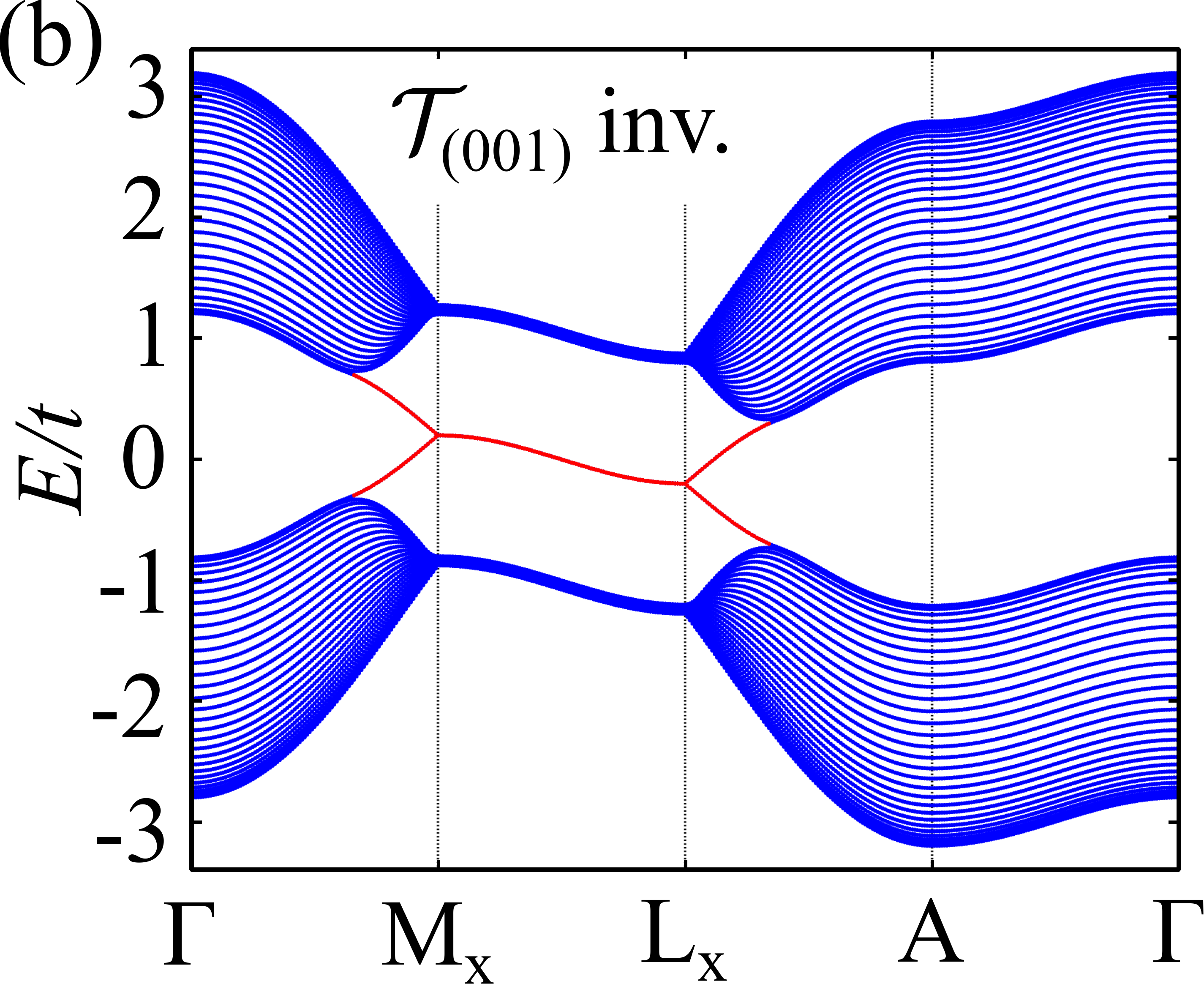}}\\
\subfloat{\includegraphics[width=0.49\linewidth]
{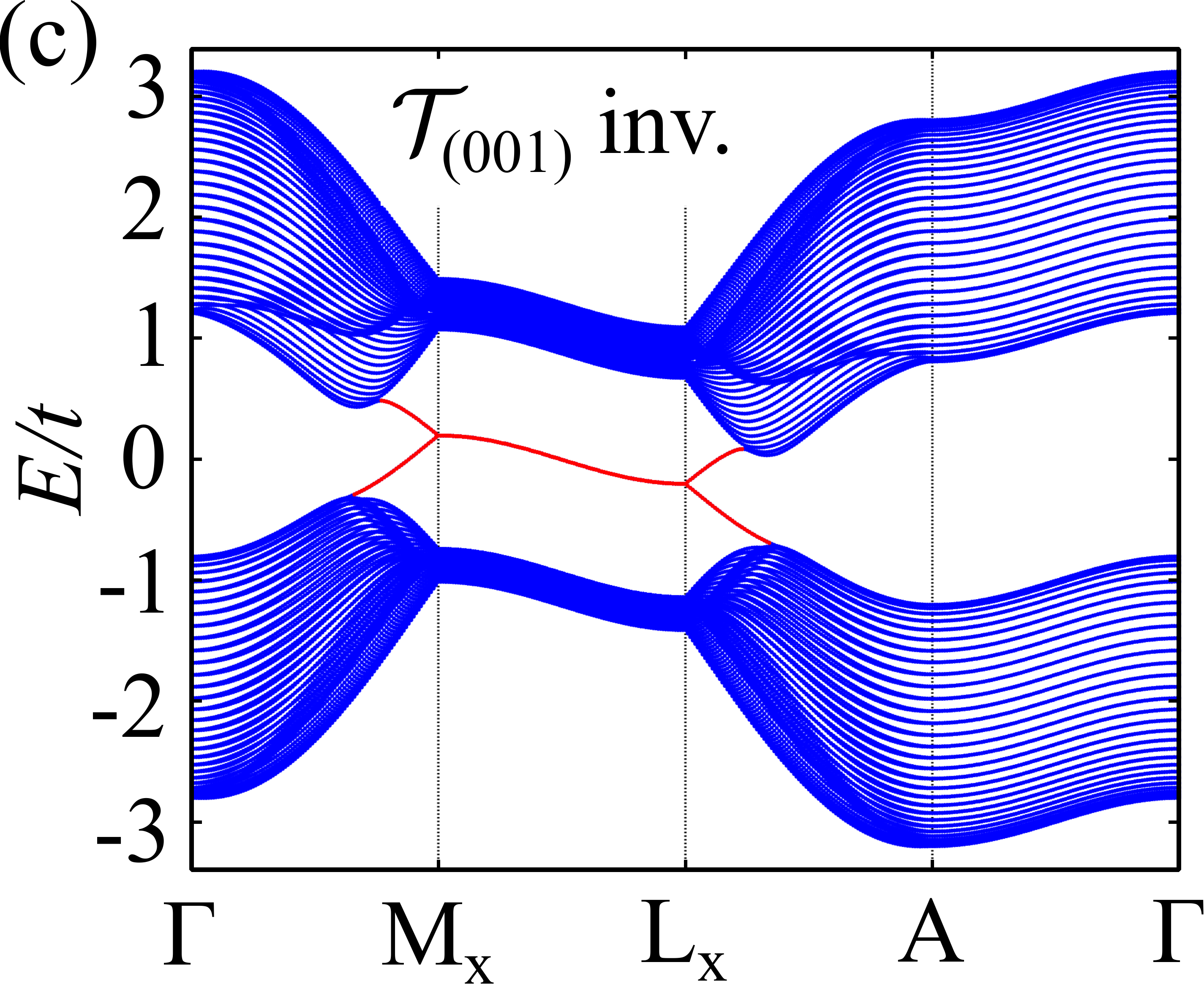}}
\hfill
\subfloat{\includegraphics[width=0.49\linewidth]
{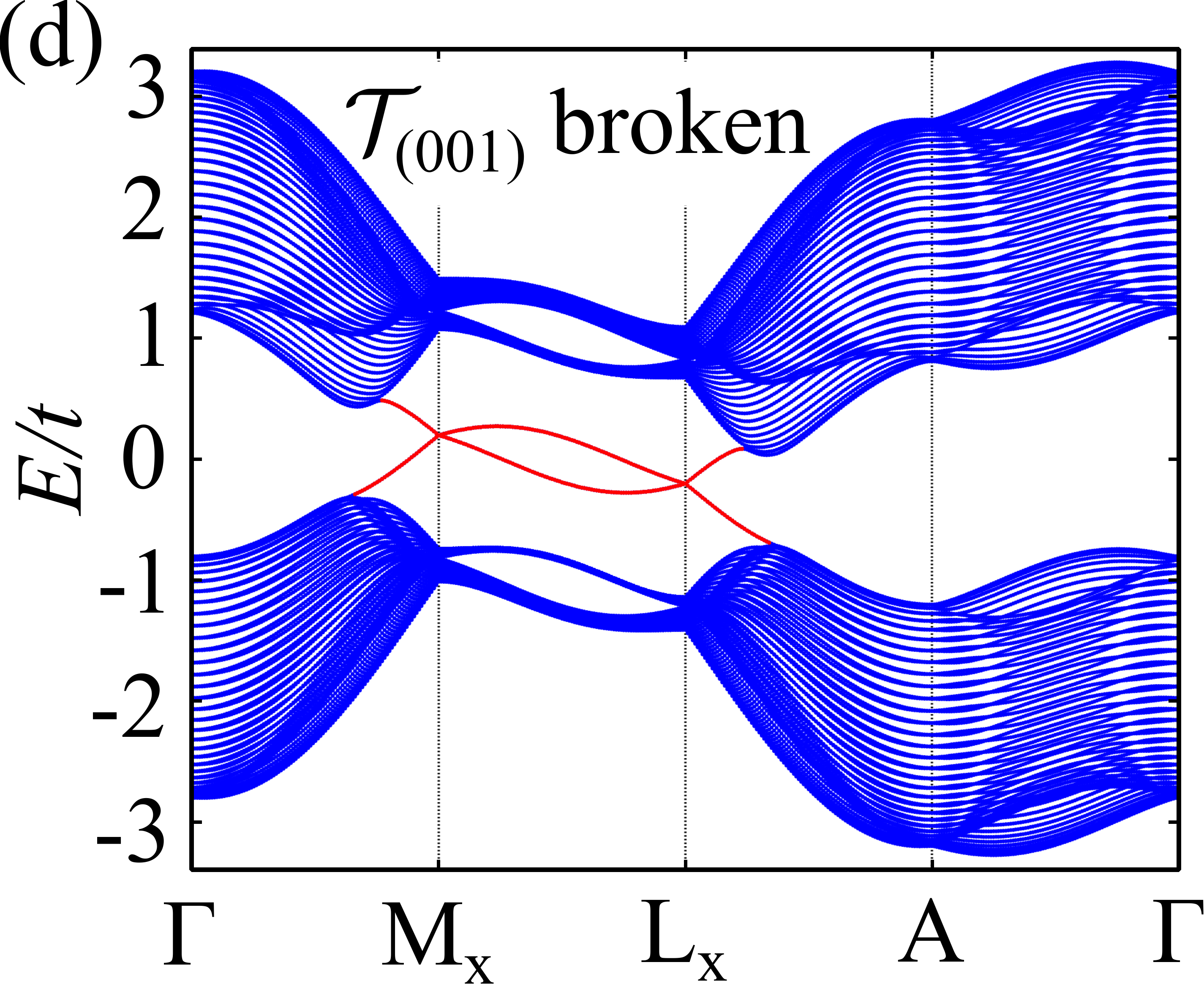}}
\caption{(color online). Energy bands for the stacked Kane-Mele model: bands
of a (010) slab of width $W=30$ are shown along high symmetry lines of
the surface BZ for different model parameters (only nonzero parameters are
listed in units of $t$):
(a) $\lambda_\mathrm{SO}=0.1$, 
(b) $\lambda_\mathrm{SO}=\tau=0.1$,
(c) $\lambda_\mathrm{SO}=\tau=\lambda_\mathrm{R}=0.1$,
(d) $\lambda_\mathrm{SO}=\tau=\lambda_\mathrm{R}
=\lambda_{\textrm{SO},\perp}=0.1$.
Surface states are highlighted in red.
Note the Dirac line in panels (a)-(c) which is split into two 
Dirac points in panel (d) due to the breaking of $\mathcal{T}_{(001)}$
symmetry.}
\label{fig:stacked_KM_bands}
\end{figure}

\begin{figure}[t]\centering
\subfloat{\includegraphics[width=0.49\linewidth]
{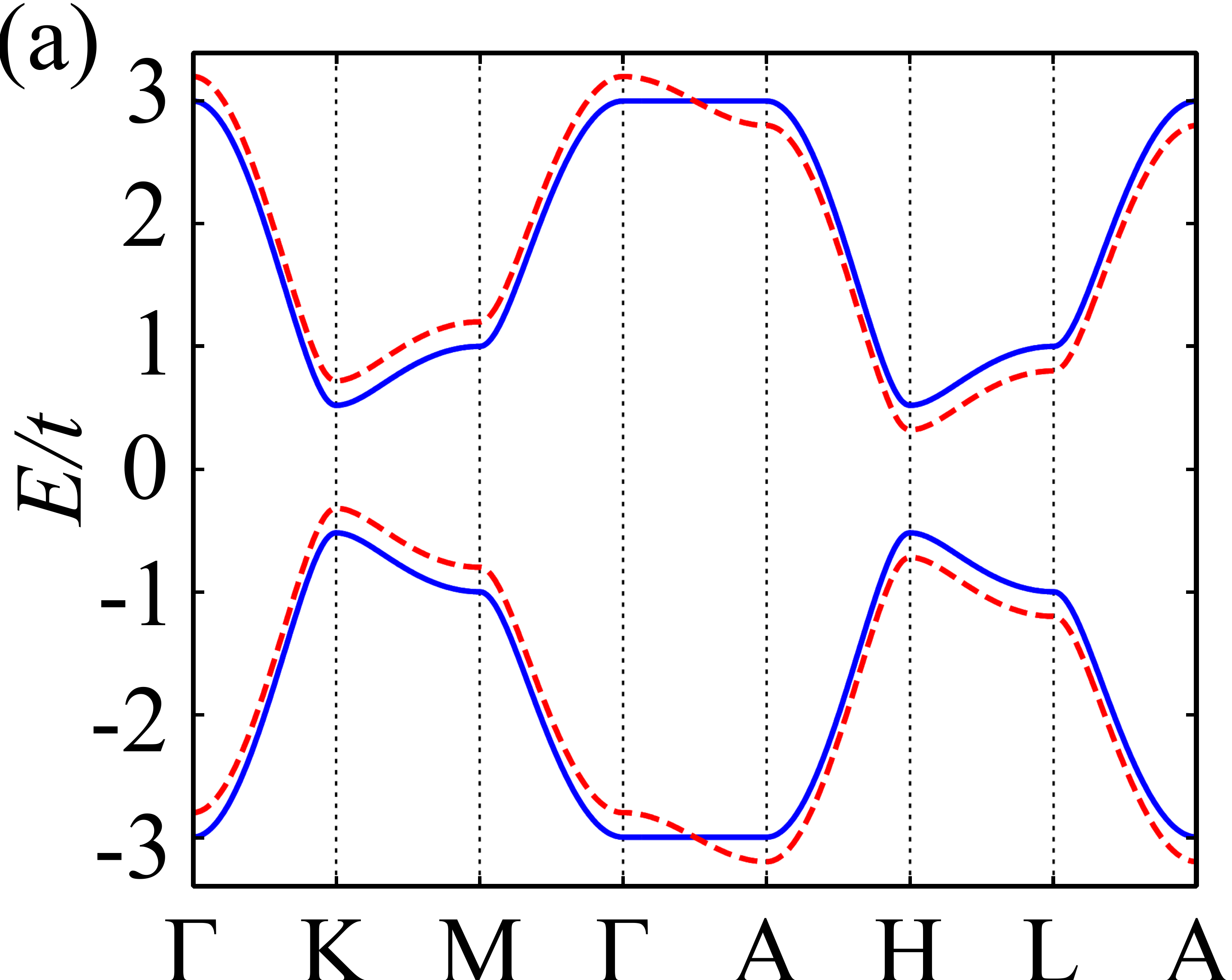}}
\hfill
\subfloat{\includegraphics[width=0.49\linewidth]
{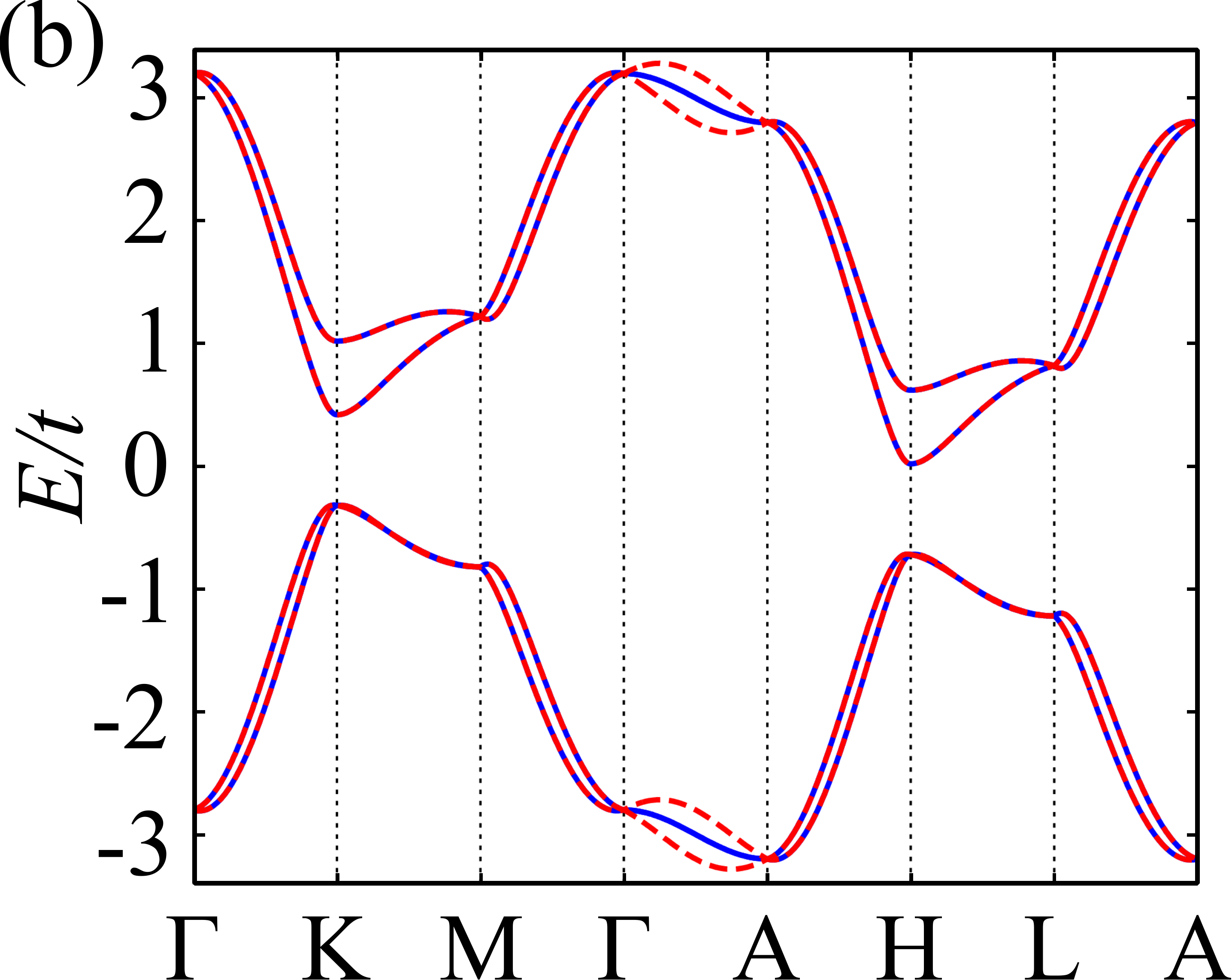}}
\caption{(color online). Bulk energy bands for the stacked Kane-Mele model
along high symmetry lines of the BZ for different model parameters 
(only nonzero parameters are listed in units of $t$):
(a) with inversion symmetry: $\lambda_\mathrm{SO}=0.1$ (solid blue lines), 
$\lambda_\mathrm{SO}=\tau=0.1$ (dashed red lines);
(b) broken inversion symmetry: $\lambda_\mathrm{SO}=\tau=\lambda_\mathrm{R}=0.1$ 
(solid blue lines), $\lambda_\mathrm{SO}=\tau=\lambda_\mathrm{R}
=\lambda_{\textrm{SO},\perp}=0.1$ (dashed red lines).}
\label{fig:stacked_KM_bands_bulk}
\end{figure}
In Fig.~\ref{fig:stacked_KM_bands}, the band structure of the (010) Kane-Mele
slab along high-symmetry lines of the surface BZ is shown for different
model parameters. Here, we ignore the sublattice potential term 
in $H_\textrm{KM}$.
The effect of this term will be discussed at the end of this section.
In addition, in Fig.~\ref{fig:stacked_KM_bands_bulk}
we also plot
the corresponding bulk energy bands along high symmetry lines of the 
bulk BZ. 

In Fig.~\ref{fig:stacked_KM_bands}(a), only the in-plane hopping and
in-plane SOC are nonzero. For the chosen parameters,
the bulk spectrum exhibits an energy gap and we find surface states 
traversing the bulk gap. The band structure along $\overline{\Gamma M_x}$ and
$\overline{L_x A}$ is identical to that of the 2D Kane-Mele model with
zigzag edges and the same parameter values (see Ref.~\onlinecite{COB13}). 
For each surface,
there is one spin-filtered surface band emerging from the bulk. The bands meet
at the TRI momenta $M_x$ and $L_x$, respectively, where we find two-fold
Kramers degeneracies, which are topologically protected by conventional
time-reversal symmetry.
Along $\overline{M_x L_x}$ we find a two-fold line
degeneracy of the topological bands -- a Dirac line. This is easily explained
in the light of in-plane time-reversal symmetry.
With the chosen parameters, the 2D Kane-Mele model 
is a topological insulator with a Dirac point at the $M_x$ point
of the surface BZ for the zigzag termination. Therefore, a stack of these
systems forms a weak TI with $\Z_2$ indices $(\nu_0;\nu_1\nu_2\nu_3)=(0;001)$.
Since $\mathcal{T}$ symmetry is preserved for the individual layers and 
the layers are not coupled, 
in-plane time-reversal symmetry $\mathcal{T}_{(001)}$ is automatically 
conserved for the
stacked system and we must find topologically protected Dirac lines. The 
flatness of the
Dirac line is due to the absence of dispersion along the $k_z$ direction. 

In Figs.~\ref{fig:stacked_KM_bands}(b-d), different terms have been added to 
the Hamiltonian of the system one after the other. In 
Fig.~\ref{fig:stacked_KM_bands}(b), interlayer hopping has been included,
which causes the band structure to disperse in the $k_z$ direction. The bands,
in particular the Dirac line, acquire a $\cos{k_z}$ dispersion since the 
interlayer hopping connects only adjacent layers. However, it is easy to check 
that
the term preserves $\mathcal{T}_{(001)}$ symmetry. Therefore, the Dirac line is 
topologically protected and we have found a fully 3D system that
exhibits a 1D Dirac particle on its surface.

It is worth mentioning that the bands
of the bulk spectrum (see Fig.~\ref{fig:stacked_KM_bands_bulk}(a)) are two-fold 
degenerate
in the entire BZ due to simultaneous conventional time-reversal and inversion
symmetry. For the surface bands of the slab, however, those symmetries 
only imply 
that corresponding topological surface bands on \emph{both} surfaces are 
degenerate.
For line degeneracies on just \emph{one} surface, in-plane time-reversal 
symmetry is 
essential as can be easily seen by adding an inversion-symmetry breaking term,
e.g. in-plane Rashba SOC [see Fig.~\ref{fig:stacked_KM_bands}(c)]. 
It preserves in-plane time-reversal symmetry but breaks inversion symmetry
as well as the remaining $\mathcal{U}(1)$ spin symmetry. Hence, in the 
bulk spectrum the two-fold degeneracies are lifted except at the Kramers points
and along the Kramers lines [see Fig.~\ref{fig:stacked_KM_bands_bulk}(b)]. 
Furthermore, in the band structure of the slab the Dirac line is not lifted 
as shown in Fig.~\ref{fig:stacked_KM_bands}(c). 
An interlayer SOC, however, breaks in-plane time
reversal symmetry, while preserving its conventional counterpart, 
and we see that the effectively
1D Dirac particle decays into two 2D Dirac particles, one at the $M_x$ point
and the other at the $L_x$ point of the surface BZ with a small shift
in energy [see Fig.~\ref{fig:stacked_KM_bands}(d)]. 

\begin{figure}[b]\centering
\subfloat{\includegraphics[width=0.49\linewidth]
{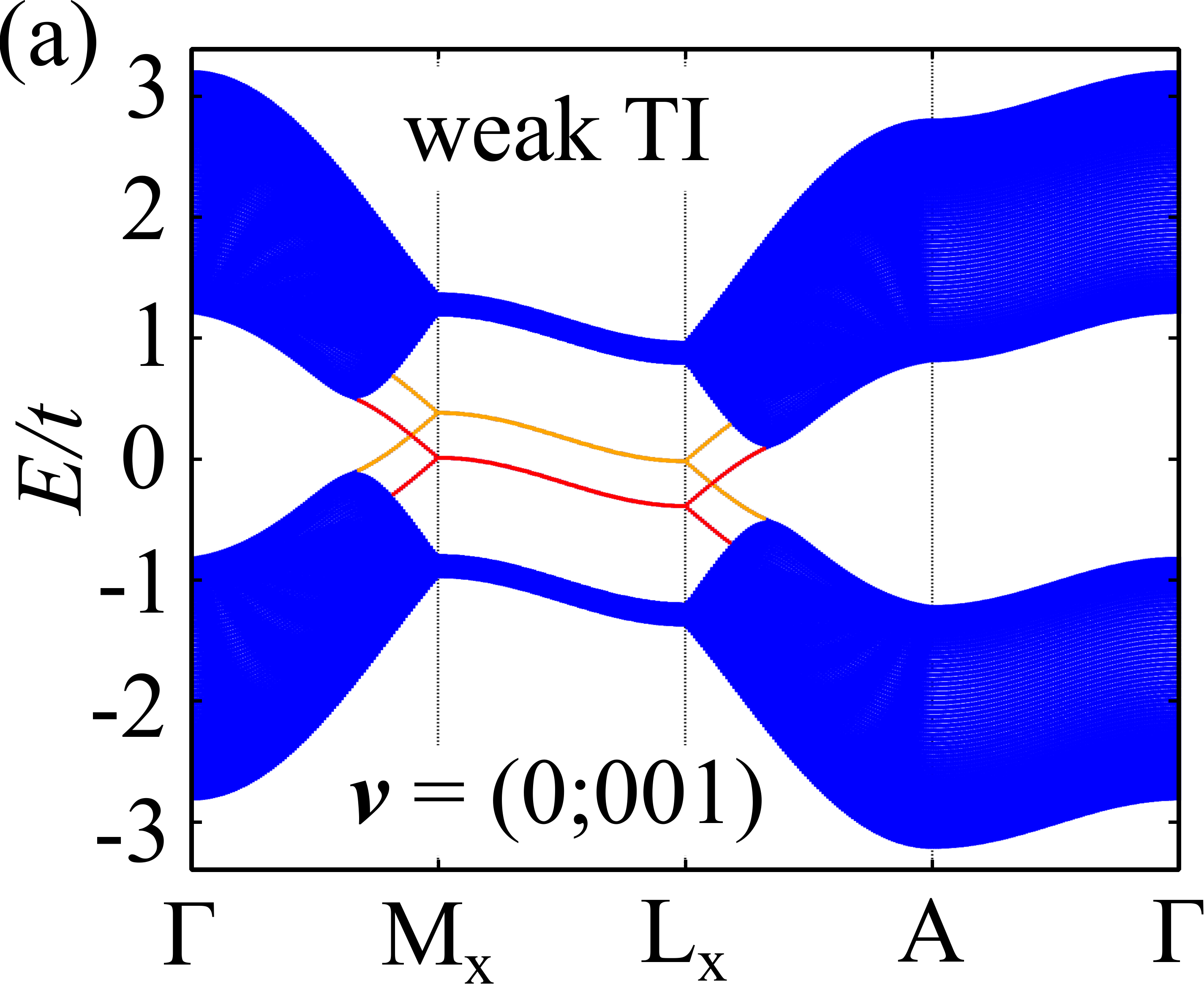}}
\hfill
\subfloat{\includegraphics[width=0.49\linewidth]
{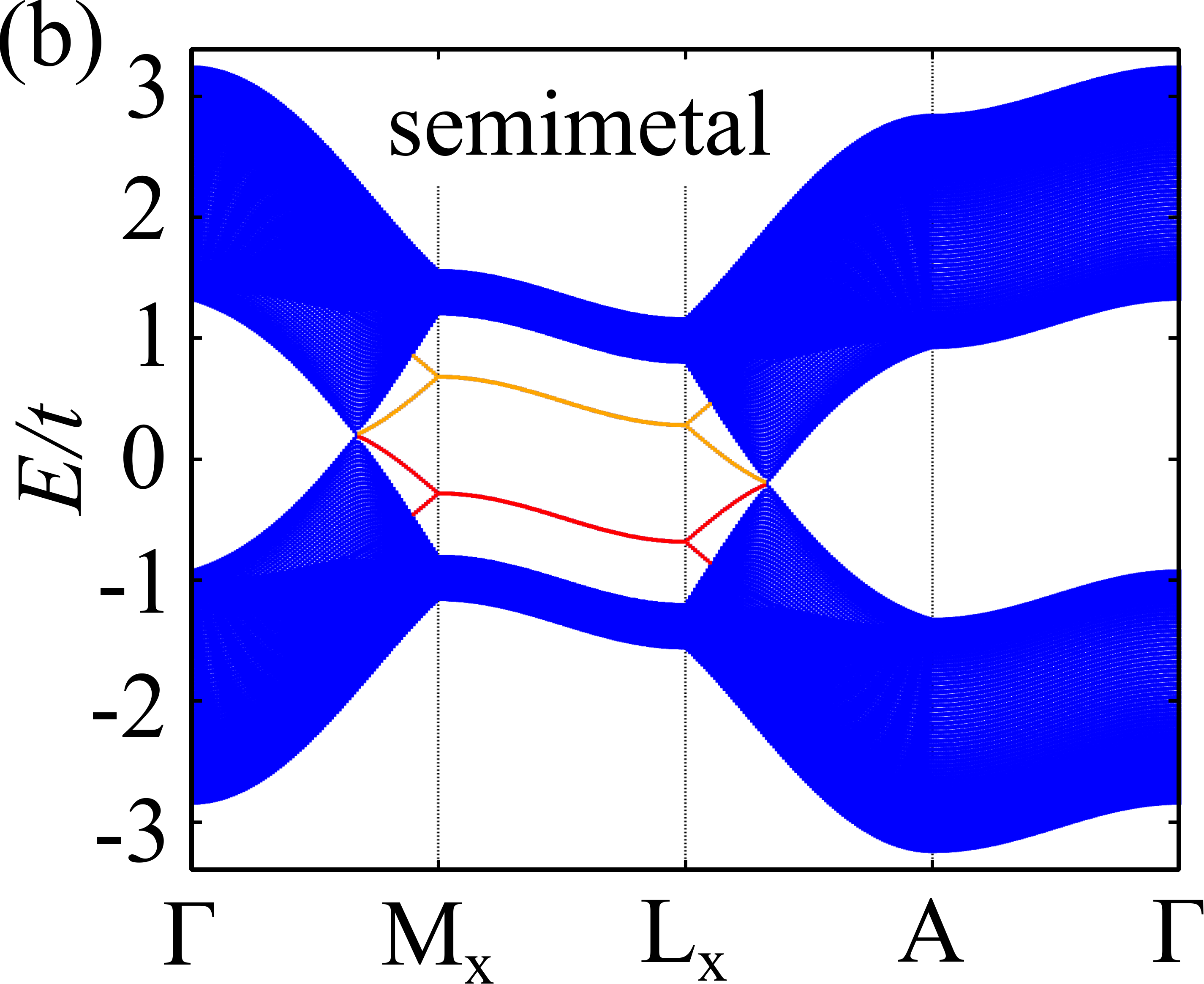}}\\
\subfloat{\includegraphics[width=0.49\linewidth]
{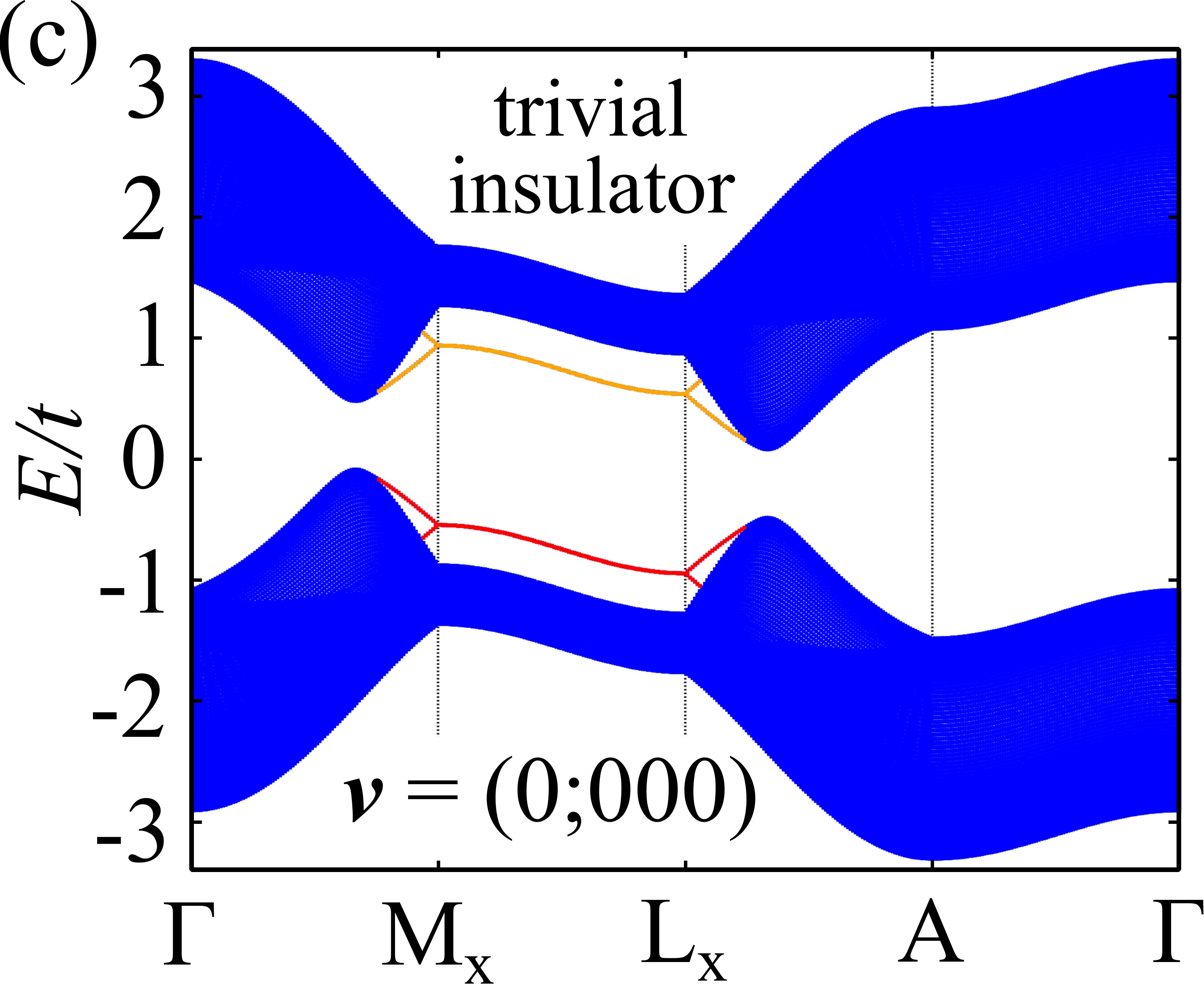}}
\caption{(color online). Energy bands for the stacked Kane-Mele model
with $\mathcal{T}_{(001)}$ invariance: bands
of a (010) slab of width $W=80$ are shown along high symmetry lines of
the surface BZ for different values of the mass term (in units of 
$t$):
(a) $\lambda_\nu=0.2$, 
(b) $\lambda_\nu=0.52$,
(c) $\lambda_\nu=0.8$.
The mass term does not break in-plane time-reversal 
invariance with respect to the $xy$ plane.
The only other nonzero model parameters are $\lambda_\mathrm{SO}=\tau=0.1t$.
Surface states are highlighted in red
(left surface) and orange (right surface).
Note the transition from a weak TI to a trivial insulator at 
$\lambda_\nu=0.52$.}
\label{fig:stacked_KM_bands_mass}
\end{figure}
So far, we have
ignored the staggered sublattice 
potential term in $H_\textrm{KM}$.
However, it is well-known 
that such a mass term can result in a transition from a topological to 
a trivial insulator
in the 2D Kane-Mele model by closing and reopening the bulk energy
gap.~\cite{KaM05_1,KaM05_2} What happens to the Dirac line in the stacked
system, if we increase the mass? First of all, it is easy to check that
the mass term preserves the relevant in-plane time-reversal symmetry.
Therefore, the Dirac line cannot be destroyed in the process. But how
can the surface states then be trivial in the trivial sector? The key is
the closing of the bulk energy gap. Before, the surface states are connected
to both the upper and the lower bulk continuum. However, the process of
closing and reopening the gap 
allows them to merge only with the upper \emph{or} the lower bulk
continuum, respectively. In this way, the Dirac line remains intact but the
surface states do no longer traverse the bulk energy gap and are, therefore,
topologically trivial (see Fig.~\ref{fig:stacked_KM_bands_mass}).


\section{Cubic Liu-Qi-Zhang model}
\label{sec:cubic_LQZ_model}

Let us now study in-plane time-reversal invariance in the context of 
another, more involved model, namely the cubic Liu-Qi-Zhang 
Hamiltonian.~\cite{LQZ12} It
is derived from a model introduced by 
Zhang \textit{et al.},~\cite{ZLQ09} which has been succesfully used to 
describe the Bi$_2$Se$_3$ family of strong topological insulators. 
It is a 3D nearest-neighbor tight-binding model on a simple
cubic lattice with two orbital and two spin degrees of freedom per site. The
corresponding Hamiltonian in momentum representation is~\cite{LQZ12}
\begin{eqnarray}
\mathcal{H} &=& \sum_\mathbf{k}\sum_{a,b=1}^2\sum_{\sigma\sigma'}
H_{a\sigma,b\sigma'}(\mathbf{k})\, 
d_{\mathbf{k}a\sigma}^\dagger d_{\mathbf{k}b\sigma'},
\label{eq:H_Liu_old_coord}
\end{eqnarray}
with the $4\times 4$ Bloch Hamiltonian~\cite{LQZ12}
\begin{eqnarray}
H(\mathbf{k}) &=& [M_0 + 6B - 2B\sum_{i=1}^3 \cos k_i]\,\Gamma_5
\nonumber\\
&&{}+ A \sum_{i=1}^3 \Gamma_i \sin k_i. 
\label{eq:Bloch_Liu_old_coord}
\end{eqnarray}
Here, $\Gamma_j$ denotes the Dirac matrices of 
Ref.~\onlinecite{LQZ12}, where also the introduced parameters and notations
are explained. We note that the Dirac matrices are Kronecker products
of Pauli matrices $s^i$ in spin space and Pauli matrices $\sigma^i$ 
in orbital space. The coordinate system used is aligned
with the edges of the cubic unit cell and we write 
$\mathbf{k}=(k_1,k_2,k_3)\equiv(k_x,k_y,k_z)$. The model describes
a trivial insulator for $M_0>0$ and $M_0<-12$, a strong TI with $\Z_2$ indices
$(1;000)$ or $(1;111)$ for $0>M_0>-4B$ or 
$-8B>M_0>-12B$, and a weak TI with $\Z_2$ indices
$(0;111)$ for $-4B>M_0>-8B$,~\cite{LQZ12} where in all cases we have $A=B$.
In particular, we focus on the weak TI phase $(0;111)$. For this set-up,
surface states are studied for a slab of thickness $W$ with $(001)$
surfaces. The corresponding Bloch Hamiltonian, we have to work with, is 
$4W\times 4W$ 
and the band structures in the slab
geometry are obtained by exact diagonalization.

\begin{figure}[t]\centering
\includegraphics[width=1.0\linewidth]
{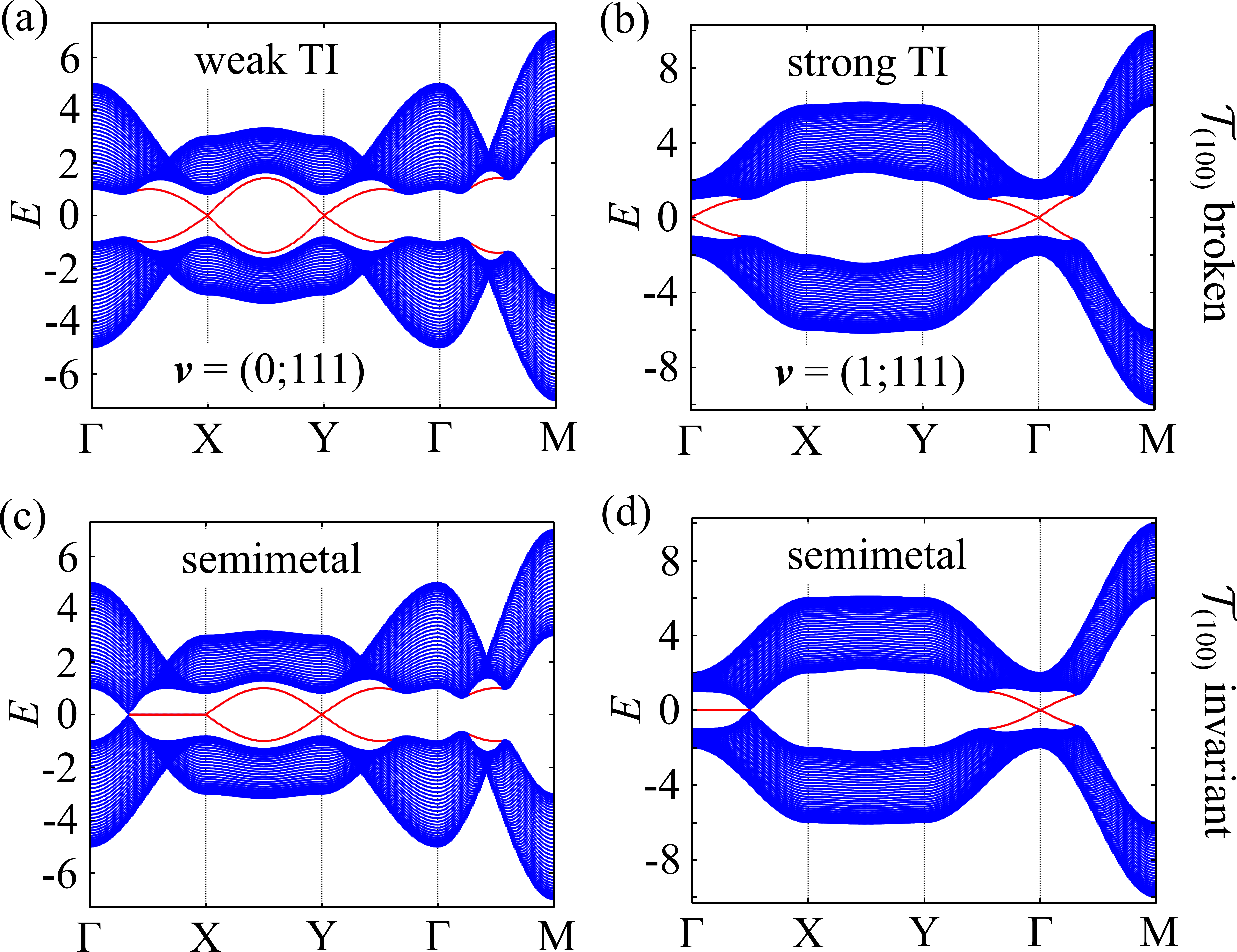}
\caption{(color online). Energy bands for the cubic Liu-Qi-Zhang model: bands
of a (001) slab of width $W=40$ are shown along high symmetry lines of
the surface BZ associated with the original cubic unit cell. The model
parameters are $M_0=-5.0$, $A=B=1.0$ for panels (a), (c), and 
$M_0=-2.0$, $A=B=1.0$ for panels (b), (d). We show the band structures
for broken (upper panels) and restored (lower panels) in-plane time-reversal 
symmetry with respect to the $yz$ plane. Surface states are highlighted in red.}
\label{fig:LQZ_bands_old_coord}
\end{figure}
Let us first check the weak TI phase for in-plane time-reversal
invariance with respect to the $yz$ plane. The corresponding operator is
\begin{equation}
\mathcal{T}_{(100)} = i(s^y \otimes \id)K\:\:
\mathrm{with}\:\: k_y,k_z\rightarrow -k_y,-k_z.
\end{equation}
It is easy to verify
that this symmetry is broken due to the $A\sin k_x\, \Gamma_1$
term in Eq.~\eqref{eq:Bloch_Liu_old_coord}. In 
Fig.~\ref{fig:LQZ_bands_old_coord}, we show what happens to the band structure
of the systems in a $(001)$ slab geometry, 
if we tune the symmetry-breaking term to zero. Obviously, the
two-fold degeneracy for the surface states along $\overline{\Gamma X}$
($\overline{YM}$ is not shown) is restored. However, along this
line the surface states were connected to the bulk before carrying out the 
deformation. For this reason, they 
pull the bulk bands down, resulting in a closing of the bulk
energy gap. Hence, the system undergoes a semimetal transition, if we
reestablish the in-plane time-reversal symmetry with respect to the $yz$
plane. Besides, this would have happened also for the $xz$ and the $xy$
plane (not shown).

\begin{figure}[t]\centering
\includegraphics[width=1.0\linewidth]
{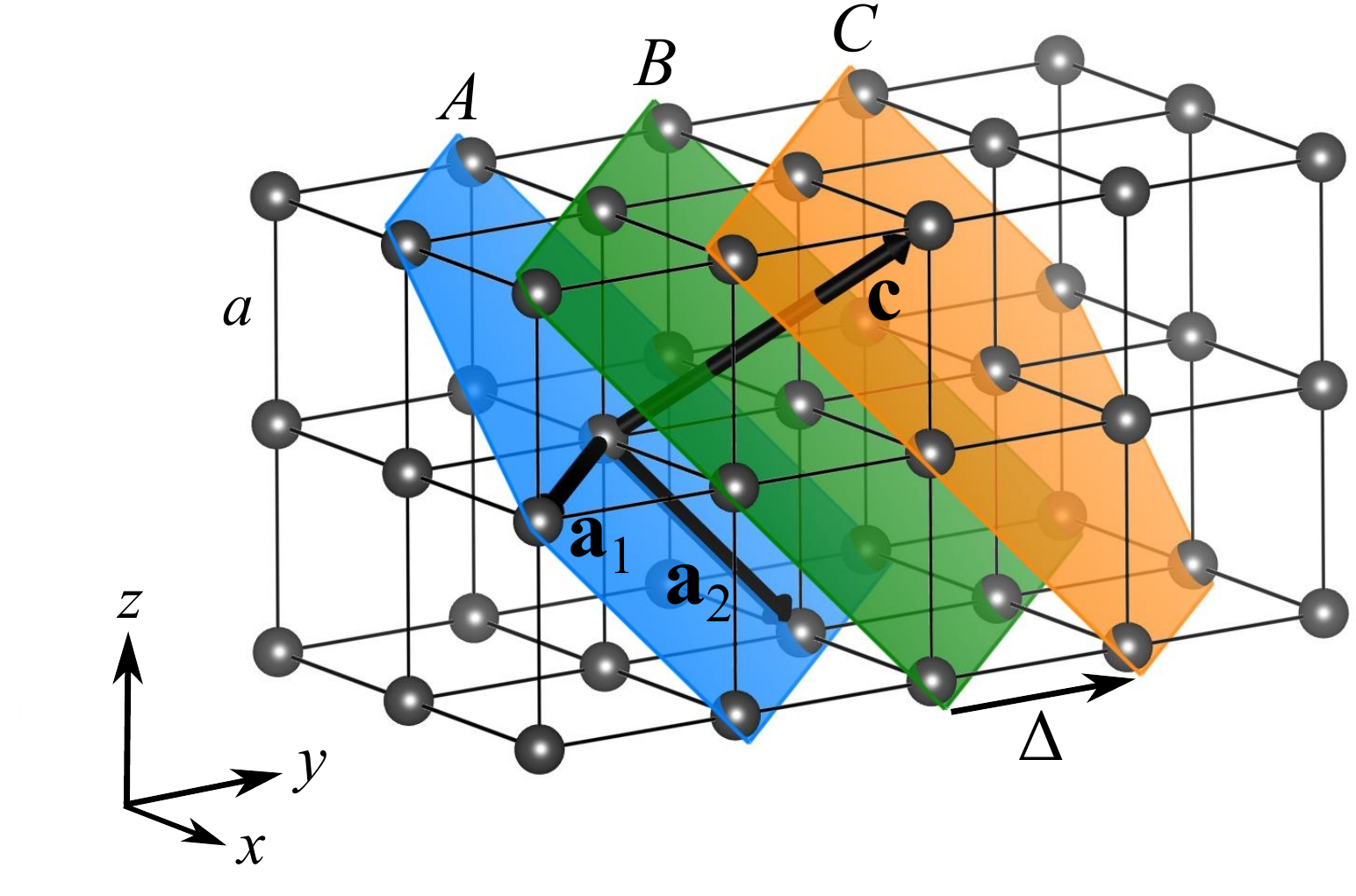}
\caption{(color online). Alternative description of the lattice
in the cubic Liu-Qi-Zhang model: the three inequivalent layers
corresponding to different elements $A,B,C$ of the new hexagonal unit cell 
are illustrated. Within the layers, the hexagonal structure of the
basal plane is clearly visible. Moreover, the new primitive lattice vectors 
$\mathbf{a}_1,\mathbf{a}_2,\mathbf{c}$ and the shift vector $\Delta$
are shown.}
\label{fig:LQZ_lattice}
\end{figure}

Nevertheless, by analogy with the stacked Kane-Mele model we expect the weak
TI with $(0;111)$ to develop Dirac lines without a semimetal transition, 
if we restore the in-plane
time-reversal symmetry with respect to the $(111)$ plane -- the plane described
by the weak indices. For convenience, we choose a different unit cell with
a different coordinate system attached to it. Since a weak TI with 
weak indices $(111)$ is topologically equivalent to a stack of 2D
TIs stacked along the $(111)$ direction, we are going to construct the 
new unit cell in this light (see Fig.~\ref{fig:LQZ_lattice}).

The $Z$ axis of the new coordinate system points along the (111) direction of
the original coordinate system. We further want to align the $c$ 
axis of the new unit cell
with the new $Z$ axis. It turns out that the cubic lattice is 
best described, in this way, by a hexagonal unit cell with a basis comprising 
three elements. The primitive lattice vectors of the new unit cell are
$\mathbf{a}_1=a(1,-1,0)$, $\mathbf{a}_2=a(0,1,-1)$, 
$\mathbf{c}=a(1,1,1)$ with respect to the original coordinate system, and 
$\mathbf{a}_1=\sqrt{2}a(1,0,0)$, $\mathbf{a}_2=\sqrt{2}a(-1,\sqrt{3},0)$, 
$\mathbf{c}=a(0,0,1)$ with respect to the new rotated coordinates. 
The elements of the
basis lie in different planes, where corresponding adjacent points
are relatively shifted by a vector 
$\Delta=1/3\,(-\mathbf{a}_1 + \mathbf{a}_2 +\mathbf{c})$.

After a Fourier transformation of \eqref{eq:H_Liu_old_coord} to real
space, the model parameters can be translated to the new coordinate system.
Another Fourier transformation back to momentum space then yields
a new Bloch Hamiltonian $\tilde{H}(\mathbf{k})$ which
is a $12\times 12$ matrix
due to the additional sublattice degrees of freedom $m,m'$. Here, $m,m'$ can
assume the values $A,B,C$.
As in Ref.~\onlinecite{LQZ12}, the spin parts and the orbital parts 
can be expanded in terms of $\Gamma$ matrices. In addition, the sublattice part
can be expanded in $3\times 3$ Gell-Mann matrices~\cite{LYF13} $\lambda_i$ 
and the
corresponding unit matrix denoted by $I$.
The components of the crystal momentum $\mathbf{k}$
are now $k_X,k_Y,k_Z$ with respect to the rotated coordinates. 
With this, the explicit structure of the Bloch Hamiltonian is
\begin{eqnarray}
\tilde{H} = \tilde{H}_0\otimes\Gamma_5
+ \frac{A}{2}\sum_{i=1}^3\tilde{H}_i \otimes \Gamma_i,
\end{eqnarray}
with
\begin{eqnarray}
\tilde{H}_0 &=& (M_0 + 6B)I - B(2\cos \tilde{X} 
+ \cos \tilde{Y})(\lambda_1+\lambda_6) \nonumber\\
&&{}+ B\sin \tilde{Y}(\lambda_2 + \lambda_7) \nonumber\\
&&{}- 2B\cos \tilde{X} \cos\tilde{Y} \cos\tilde{Z}\,\lambda_4 
- B\cos\tilde{Z}\,\lambda_4 \nonumber\\
&&{}+ 2B\cos \tilde{X} \sin\tilde{Y} \cos\tilde{Z}\,\lambda_5 
+ \underline{B\sin \tilde{Z}\,\lambda_5} \nonumber\\
&&{}- \underline{2B\cos \tilde{X} \sin\tilde{Y} \sin\tilde{Z}\, \lambda_4} 
\nonumber\\
&&{}+ \underline{2B\cos\tilde{X} \cos\tilde{Y} \sin\tilde{Z}\, \lambda_5} \\
\tilde{H}_1 &=& \sin \tilde{X} (\lambda_1+\lambda_6) 
-\cos \tilde{X} (\lambda_2+\lambda_7) \nonumber\\
&&{}+ \cos(\tilde{X}+\tilde{Y})\cos\tilde{Z}\,\lambda_5 \nonumber\\ 
&&{}+ \sin(\tilde{X}+\tilde{Y})\cos\tilde{Z}\,\lambda_4 \nonumber\\ 
&&{}- \underline{\sin(\tilde{X}+\tilde{Y})\sin\tilde{Z}\,\lambda_5} \nonumber\\ 
&&{}+ \underline{\cos(\tilde{X}+\tilde{Y})\sin\tilde{Z}\,\lambda_4}\\
\tilde{H}_2 &=& -\sin \tilde{X} (\lambda_1+\lambda_6) 
-\cos \tilde{X} (\lambda_2+\lambda_7) \nonumber\\
&&{}+ \cos(-\tilde{X}+\tilde{Y})\cos\tilde{Z}\,\lambda_5 \nonumber\\ 
&&{}+ \sin(-\tilde{X}+\tilde{Y})\cos\tilde{Z}\,\lambda_4 \nonumber\\ 
&&{}- \underline{\sin(-\tilde{X}+\tilde{Y})\sin\tilde{Z}\,\lambda_5} \nonumber\\ 
&&{}+ \underline{\cos(-\tilde{X}+\tilde{Y})\sin\tilde{Z}\,\lambda_4}\\
\tilde{H}_3 &=& \sin \tilde{X} (\lambda_1+\lambda_6) 
-\cos \tilde{X} (\lambda_2+\lambda_7) \nonumber\\
&&{}+\cos \tilde{Z}\,\lambda_5 +\underline{\sin \tilde{Z}\,\lambda_4},
\label{eq:ham_LQZ_new_coord}
\end{eqnarray}
where we have used the notations $\tilde{X}\equiv k_Xa/\sqrt{2}$,
$\tilde{Y}\equiv k_Ya\sqrt{3/2}$, and 
$\tilde{Z}\equiv k_Za\sqrt{3}$. Here, $a$ denotes the lattice constant
of the original cubic unit cell. The underlined terms break in-plane
time-reversal symmetry with respect to the $XY$ plane.

In the following, everything is expressed in terms of the new coordinate system.
The weak indices are now $(001)$. Therefore, we are particularly interested 
in the in-plane time-reversal symmetry 
with respect to the $(001)$ plane. The associated operator reads
\begin{equation}
\mathcal{T}_{(001)} = i(I \otimes s^y \otimes \id)K\:\:
\mathrm{with}\:\: k_X,k_Y\rightarrow -k_X,-k_Y
\end{equation}
where $I$ denotes the $3\times 3$ unit matrix in sublattice space
associated with the three-component basis of the lattice.

We will now study surface states for a slab of thickness $W$ with $(010)$ 
surfaces.
For this, we have to work with a $12W\times 12W$ Bloch Hamiltonian
$\tilde{H}^{(010)}(k_X,k_Z)$. The band structures in the slab
geometry are again obtained by exact diagonalization.

\begin{figure}[t]\centering
\includegraphics[width=1.0\linewidth]
{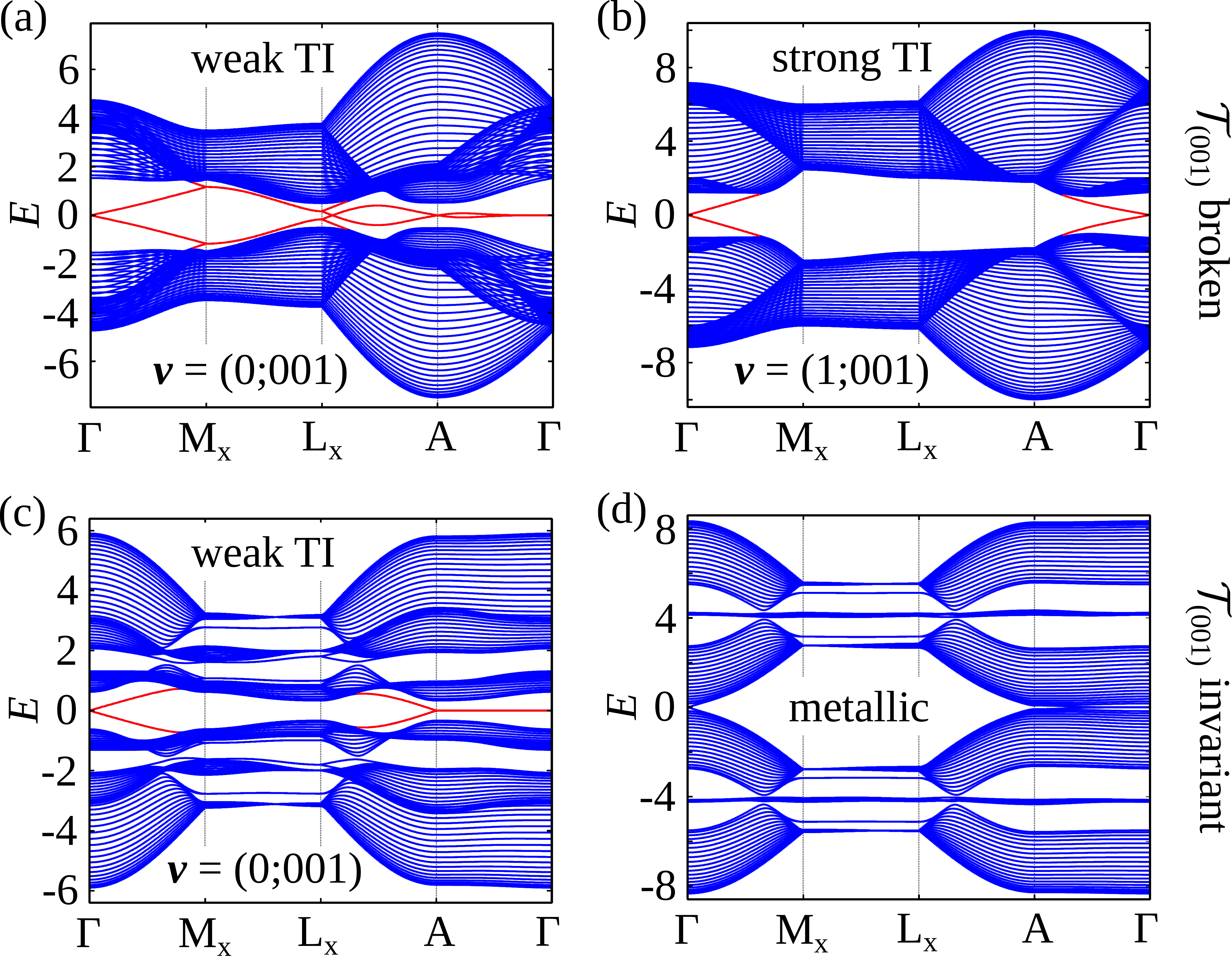}
\caption{(color online). Energy bands for the cubic Liu-Qi-Zhang model: bands
of a (010) slab (with respect to the rotated coordinate system) of width $W=20$ 
are 
shown along high symmetry lines of the surface BZ associated with the
hexagonal unit cell. The model
parameters are $M_0=-4.5$, $A=B=1.0$ for panels (a), (c), and 
$M_0=-2.0$, $A=B=1.0$ for panels (b), (d). We show the band structures
for broken (upper panels) and restored (lower panels) in-plane time-reversal 
symmetry with respect to the $XY$ plane. Relevant surface states are 
highlighted in red.}
\label{fig:LQZ_bands_new_coord}
\end{figure}
Let us first have a look at the case, where this symmetry is broken.
In Fig.~\ref{fig:LQZ_bands_new_coord}(a),
we show the band structure in a $(010)$ slab
geometry along high symmetry lines of the surface BZ associated with the 
new unit cell. We find a Dirac point at $A$, another one at $\Gamma$, and
two line degeneracies along $\overline{M_x L_x}$. However, these line
degeneracies come in pairs and could be easily pushed out 
of the bulk energy gap. For this reason,
they are trivial surface bands. This is also reflected in the fact that in-plane
time-reversal symmetry is broken and therefore, 
they are not topologically protected.

In Eq.~\eqref{eq:ham_LQZ_new_coord}, all terms of the Bloch Hamiltonian
that break in-plane time-reversal symmetry are underlined. We choose to 
tune \emph{all} $k_Z$ dependent terms to zero except the 
$\cos k_Z\, \lambda_5 \otimes \Gamma_3$ term,
which preserves in-plane TR symmetry. This is possible without closing 
the bulk energy gap, so the system stays in the weak TI phase.
The effect is shown
in Fig.~\ref{fig:LQZ_bands_new_coord}(c).
We observe that the trivial line degeneracies
along $\overline{M_x L_x}$ are pushed out of the bulk energy gap. 
Along the other trivial direction $\overline{A\Gamma}$, a Dirac line
forms, which is now topologically protected by
in-plane time-reversal symmetry.
This is in perfect agreement with the previous
observations for the stacked Kane-Mele model. 

Out of curiosity, we also ask what happens for the strong TI phase, if we 
establish in-plane time-reversal symmetry. 
This is shown in Fig.~\ref{fig:LQZ_bands_old_coord}(b),~(d) for 
$\mathcal{T}_{(100)}$ symmetry with respect to the original coordinate system,
and in 
Fig.~\ref{fig:LQZ_bands_new_coord}(b),~(d) for 
$\mathcal{T}_{(001)}$ symmetry with respect to the rotated coordinate system. 
Without the symmetry, in both cases we find one Dirac point at the $\Gamma$ 
point of the surface BZ.
However, once we restore the considered in-plane time-reversal symmetry,
the bulk
energy gap closes. As already pointed out 
at the end of Sec.~\ref{sec:in_plane_tri},
this is due to the connection of
the surface states to the bulk continuum. It causes the bulk bands to be
pulled down along $\overline{\Gamma X}$ or $\overline{A\Gamma}$, 
respectively.
Therefore, it is not possible for
strong TIs to have topologically protected Dirac lines.


\section{Conclusions}
\label{sec:conclusions}

We have shown how effectively 1D Dirac electrons appear
on the surface of weak TIs in the presence of in-plane time-reversal invariance.
One might actually view such an in-plane time-reversal invariant weak TI as
a collection of 2D QSHIs in momentum space, where the momentum component 
perpendicular to the surface described by the weak indices serves as a 
parameter.
Topologically protected 1D Dirac electrons cannot appear on the surface 
of strong TIs.
Experimentally, the surface Dirac lines connecting two time-reversal invariant 
points in an in-plane time-reversal invariant weak TI can in principle be 
detected by angle-resolved photoemission spectroscopy.

\section*{Acknowledgements}
\label{sec:acknowledgements}

We thank K. Koepernik, M. Richter, and F. Kirtschig for helpful discussions.


\begin{thebibliography}{99}

\bibitem{QiZ11}X.-L. Qi, and S.-C. Zhang, Rev.\ Mod.\ Phys.\
\textbf{83}, 1057 (2011).

\bibitem{Fu11}L. Fu, Phys.\ Rev.\ Lett.\ \textbf{106}, 106802 (2011).

\bibitem{LaT13}A. Lau, and C. Timm, Phys.\ Rev.\ B\ \textbf{88},
165402 (2013).

\bibitem{LaT14}A. Lau, and C. Timm, Phys.\ Rev.\ B\ \textbf{90},
024517 (2014).

\bibitem{AFG14}A. Alexandradinata, C. Fang, M. J. Gilbert, and A. Bernevig,
Phys.\ Rev.\ Lett\ \textbf{113}, 116403 (2014).

\bibitem{HaK10}M. Z. Hasan, and C. L. Kane, Rev.\ Mod.\ Phys.\
\textbf{82}, 3045 (2010).

\bibitem{Moo10}J. E. Moore, Nature\ \textbf{464}, 194 (2010).

\bibitem{FKM07}L. Fu, C. L. Kane, and E. J. Mele, Phys.\ Rev.\ Lett.\ 
\textbf{98}, 106803 (2007).

\bibitem{FuK07}L. Fu, and C. L. Kane, Phys.\ Rev.\ B \textbf{76}, 
045302 (2007).

\bibitem{BHZ06}B. A. Bernevig, T. L. Hughes, and S. C. Zhang, Science\ 
\textbf{314}, 1757 (2006).

\bibitem{KWB07}M. K\"onig, S. Wiedemann, C. Br\"une, A. Roth, H. Buhmann,
L. W. Molenkamp, X.-L. Qi, and S.-C. Zhang, Science\ \textbf{318}, 766 (2007).

\bibitem{HXW09}D. Hsieh, Y. Xia, L. Wray, D. Qian, A. Pal, J. H. Dil,
J. Osterwalder, F. Meier, G. Bihlmayer, C. L. Kane, Y. S. Hor, R. J. Cava,
and M. Z. Hasan, Science\ \textbf{323}, 919 (2009).

\bibitem{CAC09}Y. L. Chen, J. G. Analytis, J.-H. Chu, Z. K. Liu, S.-K. Mo,
X. L. Qi, H. J. Zhang, D. H. Lu, X. Dai, S. C. Zhang, I. R. Fisher,
Z. Hussain, and Z.-X. Shen, Science\ \textbf{325}, 178 (2009).

\bibitem{ZLQ09}H. Zhang, C. Liu, X. Qi, X. Dai, Z. Fang, and S. Zhang,
Nat.\ Phys.\ \textbf{5}, 438 (2009).

\bibitem{RIR13}B. Rasche, A. Isaeva, M. Ruck, S. Borisenko, V. Zabolotnyy,
B. B\"uchner, K. Koepernik, C. Ortix, M. Richter, and J. van den Brink,
Nat.\ Mater.\ \textbf{12}, 422 (2013).

\bibitem{LQZ12}C.-X. Liu, X.-L. Qi, and S.-C. Zhang, Physica\ E\ \textbf{44}, 
906 (2012).

\bibitem{GEY14}Q. D. Gibson, D. Evtushinsky, A. N. Yaresko, V. B. Zabolotnyy,
M. N. Ali, M. K. Fuccillo, J. Van den Brink, B. B\"uchner, R. J. Cava, 
and S. V. Borisenko, Sci.\ Rep.\ \textbf{4}, 5168 (2014).

\bibitem{KaM05_1}C. L. Kane, and E. J. Mele, Phys.\ Rev.\ Lett.\ \textbf{95}, 
226801 (2005).

\bibitem{KaM05_2}C. L. Kane, and E. J. Mele, Phys.\ Rev.\ Lett.\ \textbf{95}, 
146802 (2005).

\bibitem{FuK06}L. Fu, and C. L. Kane, Phys.\ Rev.\ B \textbf{74}, 
195312 (2006). 

\bibitem{COB13}L. Cano-Cort\'es, C. Ortix, and J. van den Brink,
Phys.\ Rev.\ Lett.\ \textbf{111}, 146801 (2013).

\bibitem{LYF13}B. Li, Z.-H. Yu, and S.-M. Fei,
Sci.\ Rep.\ \textbf{3}, 2594 (2013).

\end{thebibliography}
\end{document}